\documentclass[]{aa}
\usepackage{epsfig}
\def\lsim{\vcenter{\hbox{$<$}\offinterlineskip\hbox{$\sim$}}}
\def\gsim{\vcenter{\hbox{$>$}\offinterlineskip\hbox{$\sim$}}}
\begin{document}
\thesaurus{06(08.03.4; 08.13.2; 08.16.4; 08.23.3; 11.13.1; 13.09.6)}
\title{Mass-loss rates and dust-to-gas ratios for obscured Asymptotic Giant
       Branch stars of different metallicities\thanks{This paper is based in
       part on data obtained at the South-African Astronomical Observatory.}}
\author{Jacco Th. van Loon}
\institute{Institute of Astronomy, Madingley Road, Cambridge CB3 0HA, United
           Kingdom}
\offprints{Jacco van Loon, jacco@ast.cam.ac.uk}
\date{Received date; accepted date}
\maketitle
\markboth{Jacco Th.\ van Loon:
          mass-loss rates and dust-to-gas ratios of obscured AGB stars}
         {Jacco Th.\ van Loon:
          mass-loss rates and dust-to-gas ratios of obscured AGB stars}
\begin{abstract}

The mass-loss rates and dust-to-gas ratios of obscured Asymptotic Giant Branch
(AGB) stars are investigated for samples with different initial metallicities:
in the Small and Large Magellanic Clouds (SMC \& LMC) and in the Milky Way.
The properties of their circumstellar envelopes can be explained in a
consistent way if, both for obscured M-type AGB stars and for obscured carbon
stars, the total (gas+dust) mass-loss rate $\dot{M}$ depends only weakly on
initial metallicity whilst the dust-to-gas ratio $\psi$ depends approximately
linearly on initial metallicity.

\keywords{circumstellar matter -- Stars: mass loss -- Stars: AGB and post-AGB
-- Stars: winds, outflows -- Magellanic Clouds -- Infrared: stars}
\end{abstract}

\section{Introduction}

Asymptotic Giant Branch (AGB) stars develop strong mass loss at rates up to
$\dot{M}\sim10^{-5}$ M$_\odot$ yr$^{-1}$ or more. The optical light of AGB
stars with the highest $\dot{M}$ is almost entirely absorbed by their dusty
circumstellar envelopes (CSEs), and re-emitted at longer wavelengths. These
obscured AGB stars become very bright infrared (IR) objects, outshining any
other star in a galaxy except for a few red supergiants (RSGs). During a brief
period in their lives, they loose 50 to 80\% of their initial mass. This makes
them important contributors to the chemical enrichment of the interstellar
medium (ISM) --- they are possibly the main sources of dust particles in the
Universe.

Individual galaxies differ in their metal content as a result of different
star formation histories. For instance, two of our nearest neighbours, the
Large and Small Magellanic Clouds (LMC \& SMC) have current metallicities a
factor $\sim2$ and $\sim5$ less than the average metallicity in the Milky Way
($\sim$solar). When pursuing a quantitative description of the history of star
formation and chemical enrichment within a galaxy, it is essential to
correctly take into account the mass loss from AGB stars and RSGs. This
requires the dust-to-gas ratios in their CSEs to be known.

The paper is organised as follows: in Sect.\ 2 formulae are derived for
deriving mass-loss rates and dust-to-gas ratios from measurements of optical
depths and either luminosities or expansion velocities. Sect.\ 3 introduces
samples of obscured AGB stars in the MCs and in the Milky Way and addresses
the near-IR colours of their circumstellar envelopes. Relative mass-loss rates
and dust-to-gas ratios are determined in Sect.\ 4 for the magellanic and
galactic circumstellar envelopes around obscured AGB stars, and the results
are discussed in Sect.\ 5. In Appendix A new identifications of mass-losing
AGB stars with IRAS point sources in the LMC are presented and in Appendix B
expansion velocities are discussed.

\section{Formulae for dust-driven winds}

The optical depth $\tau$ is proportional to the column density and dust-to-gas
mass ratio $\psi$. The column density is proportional to the total mass
density $\rho$ and inner radius $R_{\rm i}$ of the dusty CSE. The dust grains
in the dust-formation zone are assumed to be in radiative equilibrium with the
incident stellar radiation field, keeping the effective temperature $T_{\rm
eff}$ and the dust condensation temperature $T_{\rm d}$ fixed. Then $R_{\rm
i}^2 \propto L$, the stellar luminosity. The continuity equation yields
$\dot{M} = 4 \pi r^2 \rho v_{\rm exp}$, with gas+dust mass-loss rate
$\dot{M}$, and expansion velocity $v_{\rm exp}$. Here the drift velocity ---
the velocity difference between the gas and dust fluids --- is neglected,
i.e.\ gas and dust are assumed to be well coupled (cf.\ Lamers \& Cassinelli
1999). Gas pressure and wind-driving mechanisms other than radiation pressure
on dust are neglected as well, although they become important for wind speeds
below a few km s$^{-1}$ (e.g.\ Steffen et al.\ 1997, 1998). It follows that
the optical depth
\begin{equation}
\tau(\lambda) \propto \frac{ \psi \dot{M} }{ v_{\rm exp} \sqrt{L} }
\end{equation}
The constant of proportionality includes a factor $\kappa(\lambda)$, the
wavelength-dependent opacity of the dust, and a temperature dependence as
$(T_{\rm d}/T_{\rm eff})^2$.

In a radiation-driven outflow the matter-momentum flux is coupled with the
stellar photon-momentum flux via the momentum equation (Gail \& Sedlmayr 1986)
\begin{equation}
\dot{M} v_{\rm exp} \propto \tau L
\end{equation}
with $\tau$ the flux-weighted optical depth. In general, the ratio of $\tau$
and $\tau(\lambda)$ depends on the mass-loss rate. Detailed computations such
as those presented in Habing et al.\ (1994), however, show that this
dependence vanishes for $\dot{M}>10^{-5}$ M$_\odot$ yr$^{-1}$, which applies
to the class of obscured AGB stars under study here (van Loon et al.\ 1999b).
Combination of Eqs.\ (1) and (2) then leads to a description of the expansion
velocity in terms of dust-to-gas ratio and luminosity:
\begin{equation}
v_{\rm exp} \propto \sqrt{\psi} \sqrt[4]{L}
\end{equation}
Eq.\ (2) can also be used to eliminate $v_{\rm exp}$ from Eq.\ (1), yielding a
relation between the mass-loss rate and dust-to-gas ratio, and the optical
depth and luminosity:
\begin{equation}
\log{\dot{M}} + 0.5 \log{\psi} + {\rm constant} = \log{\tau} + 0.75 \log{L}
\end{equation}
Alternatively, Eq.\ (2) can be used to eliminate $L$ from Eq.\ (1), yielding a
relation between the mass-loss rate and dust-to-gas ratio, and the optical
depth and expansion velocity:
\begin{equation}
\log{\dot{M}} + 2 \log{\psi} + {\rm constant} = \log{\tau} + 3 \log{v_{\rm
exp}}
\end{equation}
For magellanic stars luminosities are easier to measure than expansion
velocities, and for these stars it is advantageous to make use of Eq.\ (4).
For galactic stars the opposite is true, and for them Eq.\ (5) is the formula
to use.

The constants in Eqs.\ (4) and (5) are related. Calling the constants of
proportionality of Eqs.\ (1) and (2) respectively $\alpha$ and $\beta$, the
constant in Eq.\ (4) equals $\log(\alpha/\beta)$ and the constant in Eq.\ (5)
equals $\log(\alpha^2\beta)$. The constant of proportionality in Eq.\ (3)
equals $\sqrt{\alpha\beta}$. The values of $\alpha$ and $\beta$ depend on the
properties of the dust species, and calibrating them is an important yet
extremely difficult task that will not be exercised here.

\section{Samples of obscured AGB stars}

%
%
\begin{figure}[tb]
\centerline{\psfig{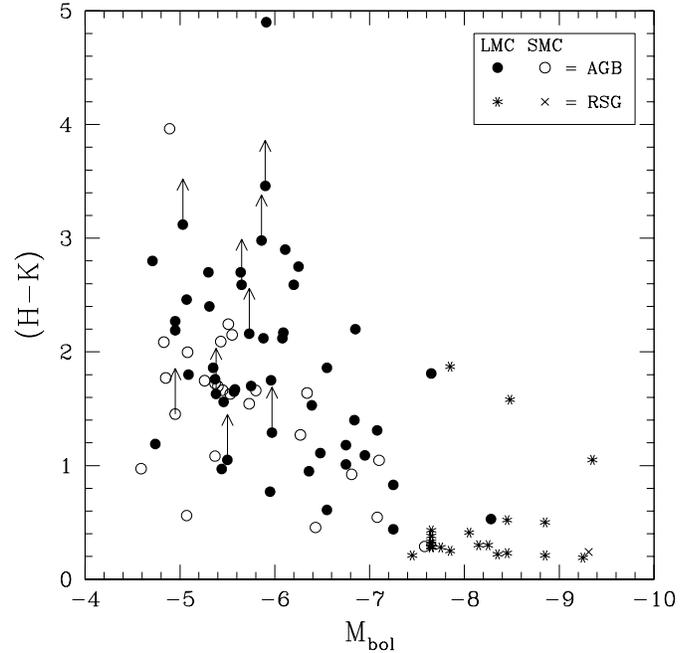}}
\caption[]{$(H-K)$ colour versus bolometric magnitude for obscured AGB stars
and red supergiants in the Magellanic Clouds. Colours are bluer for the more
luminous stars.}
\end{figure}

%
%
\begin{figure}[tb]
\centerline{\psfig{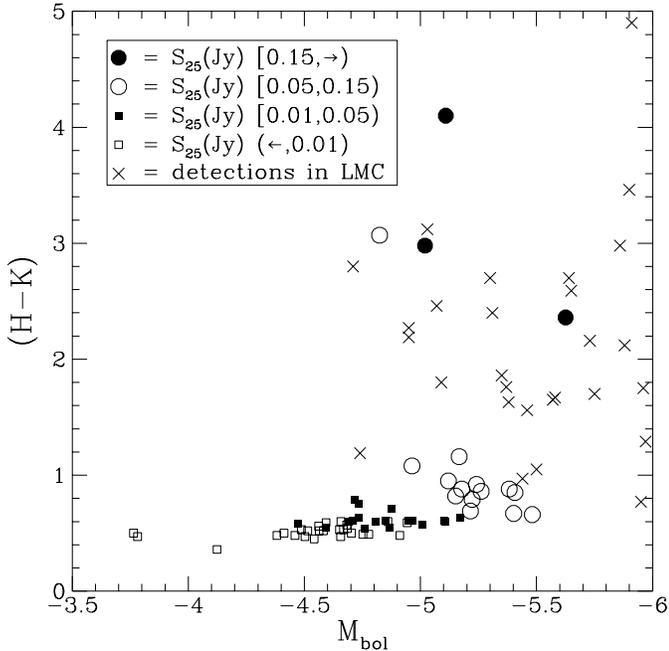}}
\caption[]{$(H-K)$ colour versus bolometric magnitude for AGB stars in the
South Galactic Cap (Whitelock et al.\ 1994). Different symbols are used
according to the 25 $\mu$m flux density measured if the star were at the
distance of the LMC. Crosses represent obscured AGB stars in the LMC (from
Fig.\ 1).}
\end{figure}

The optical depth scales with near-IR colours: $\tau \propto (H-K)-(H-K)_0$.
The stellar photosphere is assumed to have $(H-K)_0 = 0.2$ mag. Fig.\ 3 in
paper IV shows that galactic M-stars have $(H-K)_0\sim0.2$ mag, whilst
galactic carbon stars may be slightly redder with $(H-K)_0\sim0.4$ mag. The
sensitivity to the choice of $(H-K)_0$ rapidly vanishes as stars become
obscured at $(H-K)\gsim1$ mag.

Fig.\ 1 shows the $(H-K)$ colours of the obscured AGB stars and RSGs in the
MCs (Loup et al.\ 1997; Zijlstra et al.\ 1996; van Loon et al.\ 1997, 1998a:
papers I to IV). These samples are based on the identification of optical and
near-IR counterparts of point sources detected at 12 and/or 25 $\mu$m by IRAS.
The SMC photometry is mainly from Groenewegen \& Blommaert (1998). Bolometric
magnitudes for the SMC stars were determined in the same way as for the LMC
stars by spline fitting to the spectro-photometric energy distribution (see
Whitelock et al.\ 1994), adopting distance moduli of 18.55 and 18.97 mag for
the LMC and SMC, respectively (Walker 1999). Also included are two newly
identified obscured AGB stars in the LMC (Appendix A).

The reddest, optically thickest stars, with $(H-K)\gsim3$ mag, are found among
AGB stars with $M_{\rm bol}\sim-5$ to $-6$ mag. No such red objects are known
in the MCs among the brightest AGB stars with $M_{\rm bol}\sim-7$ mag, nor
amongst the RSGs. This is partly because bolometrically fainter stars have
smaller inner radii of the dusty CSE and smaller expansion velocities,
yielding larger $\tau$ (Eq.\ (1)) and redder $(H-K)$ colours at a given
mass-loss rate.

Whitelock et al.\ (1994) have searched for Long Period Variables (LPVs) in the
South Galactic Cap (SGC). Their sample consists of both optically bright and
obscured AGB stars. Their $(H-K)$ colours are plotted versus bolometric
luminosity in Fig.\ 2 (the Mira $P$-$L$ relation was applied), where different
symbols are used according to the 25 $\mu$m flux density measured if the star
were at the distance of the LMC. The obscured stars that are detected by IRAS
in the LMC have $S_{25}\gsim0.1$ Jy. Their SGC equivalents have very red
$(H-K)$ colours, though not redder than the reddest in the LMC (crosses).
Optically bright AGB stars in the SGC sample have typically $(H-K)\sim0.5$ and
$M_{\rm bol}\sim-4$ to $-5$ mag. The MC samples do not contain such objects
because their mid-IR emission is too faint to have been detected by IRAS at
the distances of the MCs. On the other hand, the SGC sample is devoid of the
brightest AGB stars with $M_{\rm bol}\sim-7$ mag as well as RSGs, because in
the Milky Way such massive stars are preferentially found in the galactic
plane.

There is a cluster of SGC stars with $(H-K)\sim0.8$ and $M_{\rm bol}\sim-5.2$
mag (Fig.\ 2). These stars clearly show circumstellar reddening, but the
mid-IR emission from their CSEs is just below the detection limit of IRAS when
placed at the distance of the LMC. This leaves open the possibility of the
existence in the LMC of a potentially large population of AGB stars with
moderate mass-loss rates and luminosities. Indeed, in paper III several field
stars were found that are not related to a nearby IRAS source but that
nevertheless had near-IR colours indicative of reddening. Recent ISO
observations confirm the presence of this AGB population (Loup et al.\ 1999).

Wood et al.\ (1998) find LPVs in the Galactic Centre with K-band magnitudes
from 5 to 13 after correction for interstellar extinction. At the distance of
the LMC this would yield K-band magnitudes from 9 to 17, i.e.\ within the
sensitivity of the searches in papers II \& III and in Groenewegen \&
Blommaert (1998). The interstellar extinction corrected $(H-K)$ colours of the
Galactic Centre LPVs average $\sim2$ mag and are $\sim4$ mag maximum, similar
to the $(H-K)$ colours of the obscured AGB stars in the LMC. Wood et al.\
argue that their sample includes stars with initial metallicities a few times
solar, and Blommaert et al.\ (1998) indeed find very red objects with
$(H-K)>4$ mag as inferred from their K- and L-band photometry.

Groenewegen et al.\ (1998) observed and modelled obscured carbon stars in the
Milky Way. They used the $P$-$L$ relation for carbon Miras to infer distances
to the individual stars, that are found to be typically within 2 kpc from the
Sun. Their two most obscured carbon stars have $(H-K)=5.5$ and 8.0 mag. They
also compiled a sample of oxygen-rich stars with near-IR photometry, pulsation
periods and expansion velocities, without overlap with the SGC and Galactic
Centre samples. The most obscured of these M stars have $(H-K)\sim6$ mag.

\section{Relative mass-loss rates and dust-to-gas ratios}

In determining mass-loss rates and dust-to-gas ratios in the MCs one can rely
on the use of luminosities and eliminate $v_{\rm exp}$ from Eq.\ (1).
Luminosities are difficult to measure for stars in the Milky Way, however, due
to unknown distances and severe interstellar extinction. Often the Mira
$P$-$L$ relation is applied to infer distances to individual stars, but this
relation possibly breaks down for stars whose stellar mantles have been
significantly reduced due to mass loss (e.g.\ Blommaert et al.\ 1998; Wood et
al.\ 1998; Wood 1998), or obscured AGB stars may fall on other, parallel
sequences (Bedding \& Zijlstra 1998; Wood 1999). For the Milky Way stars, $L$
is eliminated from Eq.\ (1), leaving $v_{\rm exp}$ to be measured. Because
$v_{\rm exp}$ has been measured for only a few LMC stars, Eq.\ (3) will be
used for the remaining LMC stars to estimate $v_{\rm exp}$ from $L$ (see
Appendix B). Although these estimates are likely to be accurate within a few
km s$^{-1}$, the contribution of the uncertainty in $v_{\rm exp}$ is enlarged
by a factor three in Eq.\ (5).

Eqs.\ (4) \& (5) are used here to compare the combination of mass-loss rate
and dust-to-gas ratio between the MCs and between the LMC and the Milky Way,
respectively. If the mass-loss rate depends on luminosity but not on initial
metallicity and if the luminosity distributions of the stars in these samples
are identical --- implying identical star formation histories --- then the
distributions over the combination of $\dot{M}$ and $\psi$ are identical
except for possible offsets due to different mean values for $\psi$ among the
different samples. The stars within each of the samples are likely to cover a
range in initial metallicities due to their different progenitor masses and
hence different formation epochs. Similar star formation histories, however,
are anticipated to result in similar distributions over initial metallicity,
and it is therefore meaningful to assign a mean initial metallicity and a mean
value for $\psi$ to each of the samples.

%
%
\begin{figure}[tb]
\centerline{\psfig{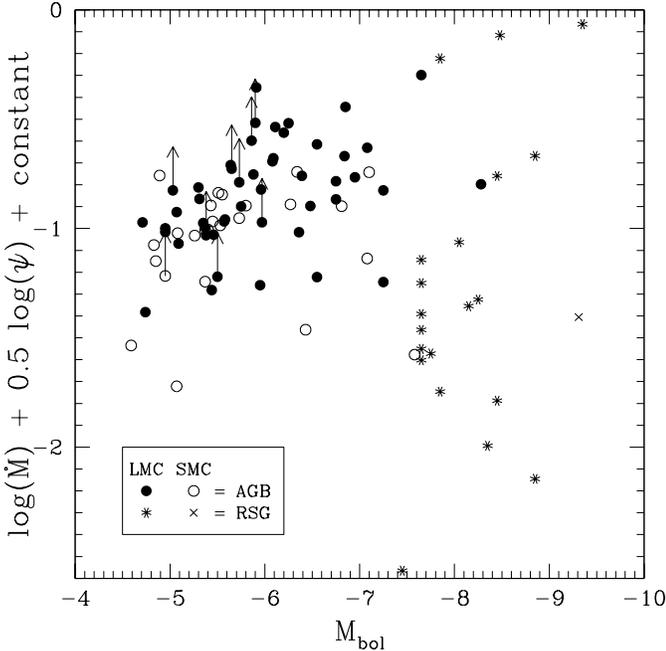}}
\caption[]{Mass-loss rate $\dot{M}$ and dust-to-gas ratio $\psi$ calculated
from Eq.\ (4), versus $M_{\rm bol}$. There is a maximum to $\dot{M}$,
increasing with higher luminosity. Mass-loss rates may be the same in SMC and
LMC if $\psi$ is sufficiently lower in the SMC.}
\end{figure}

%
%
\begin{figure}[tb]
\centerline{\psfig{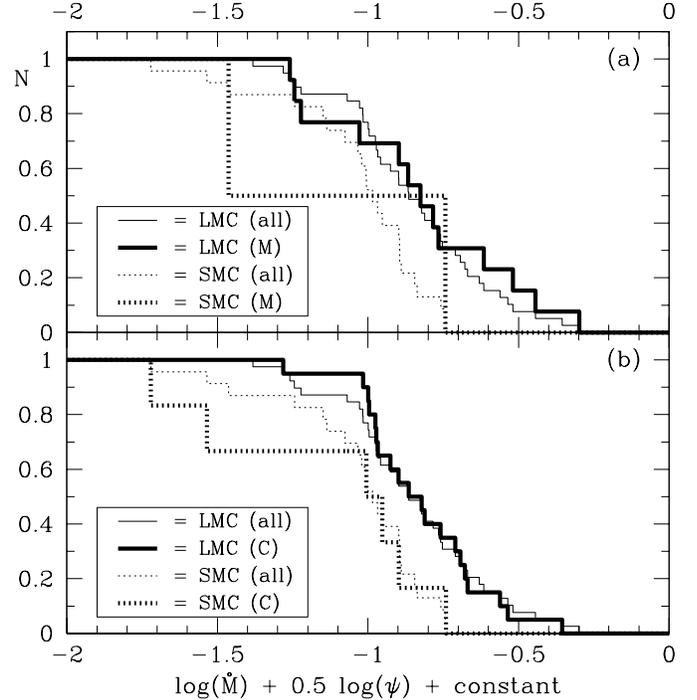}}
\caption[]{Normalised distribution of obscured AGB stars ($M_{\rm
bol}{\gsim}-7.5$ mag) over mass-loss rate $\dot{M}$ and dust-to-gas ratio
$\psi$. Boldfaced are the sub-distributions of confirmed M-type stars (a) and
confirmed carbon stars (b) amongst the obscured AGB stars.}
\end{figure}

%
%
\begin{table*}
\caption[]{Mean and error in the mean of the values of $(\log{\dot{M}} + 0.5
\log{\psi} + {\rm constant})$ for the distributions of obscured AGB stars in
the LMC and SMC. Also the differences between the various (sub-)distributions
are listed.}
\begin{tabular}{lllllll}
\hline\hline
                                                &
$\left[{\rm total}\right]$                      &
$\left[{\rm C}\right]$                          &
$\left[{\rm M}\right]$                          &
$\left[{\rm C}\right]-\left[{\rm total}\right]$ &
$\left[{\rm M}\right]-\left[{\rm total}\right]$ &
$\left[{\rm C}\right]-\left[{\rm M}\right]$     \\
\hline
$\left[{\rm LMC}\right]$                        &
\llap{$-$}$0.85\pm0.04$                         &
\llap{$-$}$0.83\pm0.05$                         &
\llap{$-$}$0.83\pm0.09$                         &
          $0.02\pm0.06$                         &
          $0.02\pm0.09$                         &
          $0.00\pm0.10$                         \\
$\left[{\rm SMC}\right]$                        &
\llap{$-$}$1.04\pm0.05$                         &
\llap{$-$}$1.14\pm0.16$                         &
\llap{$-$}$1.10\pm0.36$                         &
\llap{$-$}$0.10\pm0.17$                         &
\llap{$-$}$0.06\pm0.36$                         &
\llap{$-$}$0.04\pm0.39$                         \\
$\left[{\rm LMC}\right]-\left[{\rm SMC}\right]$ &
          $0.19\pm0.07$                         &
          $0.32\pm0.17$                         &
          $0.28\pm0.37$                         &
                                                &
                                                &
                                                \\
\hline
\end{tabular}
\end{table*}

Eq.\ (4) is used to calculate the combination of mass-loss rate and
dust-to-gas ratio from the optical depths and luminosities for the magellanic
stars (Fig.\ 3). The stars with $M_{\rm bol}<-7.5$ and $>-7.5$ mag are
considered to be RSGs and obscured AGB stars, respectively (see van Loon et
al.\ 1999b), except for the well-studied luminous obscured AGB star
IRAS05298$-$6957 ($M_{\rm bol}=-7.65$ mag) and the weakly mass-losing RSG
HV2700 ($M_{\rm bol}=-7.45$ mag). IRAS04498$-$6842 ($M_{\rm bol}=-8.28$ mag)
was classified as an AGB star (paper II), but here it is re-classified as a
RSG (see also paper IV). Fig.\ 4 shows the cumulative distributions
(normalised to unity) of the obscured AGB stars in the LMC (solid) and SMC
(dotted) over the value of $(\log{\dot{M}} + 0.5 \log{\psi} + {\rm
constant})$. These do not include the few stars that have lower limits to
their $(H-K)$ colours. The distributions for the subsamples of
spectroscopically confirmed carbon stars and oxygen-rich (M-type) stars (van
Loon et al.\ 1999b, and references therein; Groenewegen \& Blommaert 1998) are
boldfaced in Figs.\ 4a and b, respectively. Despite the small numbers for some
of the sub-distributions, especially the M-type stars in the SMC, the shapes
of the cumulative distributions are very similar. Both in the LMC and SMC the
obscured M-type AGB stars and the obscured carbon stars have distributions
that are coincident to a very high degree (Table 1). The only difference in
the distributions is the systematic offset between those in the LMC compared
to those in the SMC (Table 1). This offset is a $3 \sigma$ result for the
obscured AGB stars as a whole, a $2 \sigma$ result for the obscured carbon
stars, and consistent for the obscured M-type AGB stars (though not
significant by itself).

Eq.\ (5) is used to calculate the combination of mass-loss rate and
dust-to-gas ratio from the optical depths and expansion velocities for the
obscured AGB stars in the LMC and the Milky Way (Fig.\ 5). Only LMC stars for
which periods are known from the literature (Wood et al.\ 1992; Wood 1998) are
plotted. Stars in the different samples are not similarly distributed over
$P$, with the LMC sample having an average pulsation period longer than that
of the Milky Way sample, possibly due to selection effects or differences in
the star formation histories. Hence the populations of stars in these samples
may not be directly compared.

Two regimes of pulsation period will be focussed on: one for $500{\leq}P<800$
d, and another for $P\geq800$ d. The first group consists of stars that evolve
beyond the optically bright Mira phase with moderate $\dot{M}$ and into the
obscured AGB phase (Jura 1986). The distribution of these stars in Fig.\ 5
suggests that they still increase in mass-loss rate and/or that they represent
a large range in stellar masses. The second group consists of more evolved (or
more massive) stars for which $\dot{M}$ has become so high that they shed
their mantles on a timescale much shorter than the nuclear burning timescale.
Hence these stars stay at constant luminosity while their pulsation periods
keep increasing as their mantles are steadily diminished. The lack of any
clear correlation between $\dot{M}$ and $P$ in this part of Fig.\ 5 suggests
that thereby the mass-loss rate remains essentially constant. Stars in the LMC
sample are predominantly found in the second class of objects, whilst these
kind of stars are very rare in the Galactic Centre sample due to their
faintness in the K-band. M-type stars in the ``Solar neighbourhood'' sample of
Groenewegen et al.\ (1998) populate both regimes rather evenly.

%
%
\begin{figure}[tb]
\centerline{\psfig{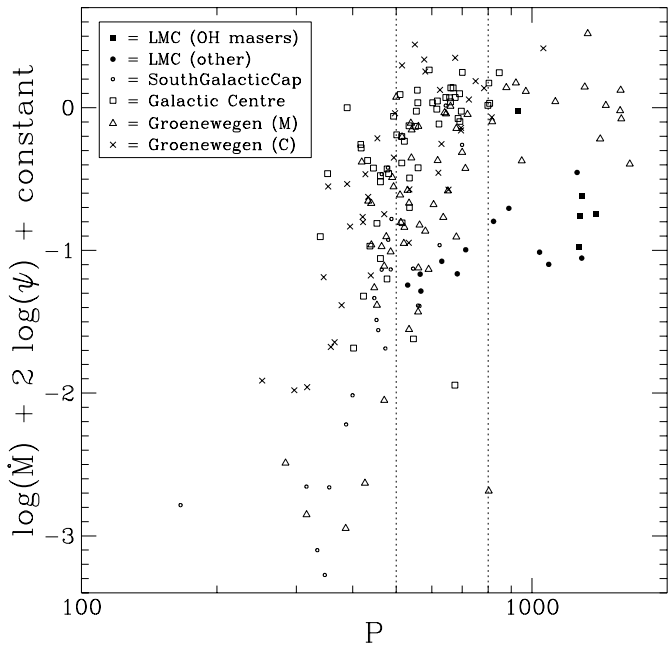}}
\caption[]{Mass-loss rate $\dot{M}$ and dust-to-gas ratio $\psi$ calculated
from Eq.\ (5), plotted versus pulsation period $P$ for obscured AGB stars in
the LMC (Wood et al.\ 1992; Wood 1998; van Loon et al.\ 1998b), South Galactic
Cap (Whitelock et al.\ 1994), Galactic Centre (Wood et al.\ 1998) and ``Solar
neighbourhood'' (Groenewegen et al.\ 1998). Dotted vertical lines delineate
regimes of periods $P<500$, $500{\leq}P<800$ and \mbox{$P\geq800$ d.}}
\end{figure}

%
%
\begin{figure}[tb]
\centerline{\psfig{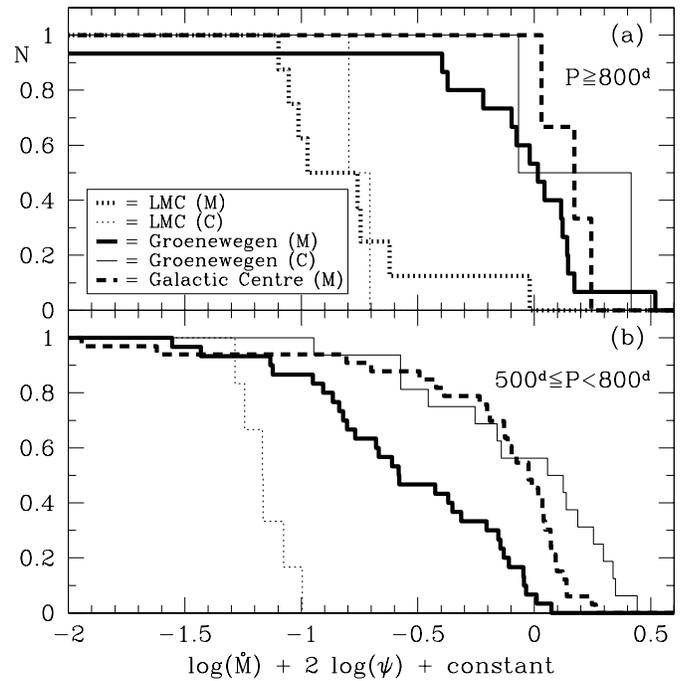}}
\caption[]{Normalised distribution of obscured AGB stars in the LMC and in the
Milky Way over mass-loss rate $\dot{M}$ and dust-to-gas ratio $\psi$ for
pulsation periods $P\geq800$ d (a) and $500{\leq}P<800$ d (b).}
\end{figure}

%
%
\begin{table}
\caption[]{Mean and error in the mean of the values of $(\log{\dot{M}} + 2
\log{\psi} + {\rm constant})$ for the distributions of obscured AGB stars in
the LMC, in the Galactic Centre (GC), and in the ``Solar neighbourhood''
(Groen; excluding the point at $P=802$ d and a value of $-2.7$). Also the
differences between the various (sub-)distributions are listed.}
\begin{tabular}{llll}
\hline\hline
                                                  &
$\left[{\rm C}\right]$                            &
$\left[{\rm M}\right]$                            &
$\left[{\rm C}\right]-\left[{\rm M}\right]$       \\
\hline
\multicolumn{4}{c}{\it $500{\leq}P<800$ d}        \\
$\left[{\rm LMC}\right]$                          &
\llap{$-$}$1.16\pm0.04$                           &
                                                  &
                                                  \\
$\left[{\rm Groen}\right]$                        &
\llap{$-$}$0.06\pm0.10$                           &
\llap{$-$}$0.55\pm0.08$                           &
          $0.49\pm0.13$                           \\
$\left[{\rm GC}\right]$                           &
                                                  &
\llap{$-$}$0.19\pm0.08$                           &
                                                  \\
$\left[{\rm Groen}\right]-\left[{\rm LMC}\right]$ &
          $1.10\pm0.11$                           &
                                                  &
                                                  \\
$\left[{\rm GC}\right]-\left[{\rm Groen}\right]$  &
                                                  &
          $0.37\pm0.12$                           &
                                                  \\
\hline
\multicolumn{4}{c}{\it $P\geq800$ d}              \\
$\left[{\rm LMC}\right]$                          &
\llap{$-$}$0.75\pm0.05$                           &
\llap{$-$}$0.79\pm0.12$                           &
          $0.04\pm0.13$                           \\
$\left[{\rm Groen}\right]$                        &
          $0.17\pm0.24$                           &
          $0.01\pm0.06$                           &
          $0.17\pm0.25$                           \\
$\left[{\rm GC}\right]$                           &
                                                  &
          $0.15\pm0.06$                           &
                                                  \\
$\left[{\rm Groen}\right]-\left[{\rm LMC}\right]$ &
          $0.93\pm0.25$                           &
          $0.79\pm0.14$                           &
                                                  \\
$\left[{\rm GC}\right]-\left[{\rm LMC}\right]$    &
                                                  &
          $0.93\pm0.14$                           &
                                                  \\
$\left[{\rm GC}\right]-\left[{\rm Groen}\right]$  &
                                                  &
          $0.14\pm0.09$                           &
                                                  \\
\hline
\end{tabular}
\end{table}

The relative distributions of mass-loss rates and dust-to-gas ratios are
derived from Fig.\ 6 for the LMC (dotted), ``Solar neighbourhood'' (M-type;
solid) and Galactic Centre (dashed), in the same way as before by estimating
the relative offsets in $(\log{\dot{M}} + 2 \log{\psi} + {\rm constant})$ of
the normalised cumulative distributions. The sample of obscured AGB stars with
$500{\leq}P<800$ d in the ``Solar neighbourhood'' (Groenewegen et al.\ 1998)
is clearly inhomogeneous: (i) the distribution of obscured M-type AGB stars is
much flatter than all other distributions, indicating a wide range of stellar
parameters, and (ii) the distribution of obscured carbon stars is offset with
respect to the distribution of obscured M-type AGB stars --- a $4 \sigma$
result (Table 2). This may reflect differences in age (initial stellar mass)
and/or initial metallicity, as the sub-samples of obscured M-type AGB stars
and obscured carbon stars in the Groenewegen et al.\ sample have been selected
in a very different way. The distributions of the obscured M-type AGB stars
and obscured carbon stars in the LMC sample are, again, indistinguishable
(Table 2). The Galactic Centre sample is positively offset with respect to
both the Groenewegen et al.\ sample ($3 \sigma$ result for $500{\leq}P<800$ d)
and the LMC sample ($7 \sigma$ result).

\section{Discussion}

\subsection{Metallicities of obscured AGB stars in the Magellanic Clouds and
in the Milky Way}

What are the initial metallicities of the obscured AGB stars in the samples
under study? First I discuss the dust-to-gas ratios in the ISM, before
discussing metallicities of relatively young stellar populations. These
considerations lead to estimates of the typical initial metallicities of the
obscured AGB stars under study. These will then be used to correlate the
mass-loss rates and dust-to-gas ratios with initial metallicity.

Most of the mass of stars that eventually evolve into AGB stars is returned to
the ISM, enriched with dust. These intermediate-mass stars are sufficiently
numerous and short-lived that the ISM has been recycled by them at least
several times over the history of their parent galaxy. A 5 M$_\odot$ star has
a lifetime of $\sim1.1 \times 10^8$ yr (Marigo et al.\ 1999), and there have
been already $\sim10^2$ generations of these massive AGB stars at work in
mixing dust with the ISM. Dust-to-gas ratios in the ISM of the LMC are found
to be $\sim\frac{1}{4}$ of that of the ISM in the Solar neighbourhood (van
Genderen 1970; Koornneef 1982; Clayton \& Martin 1985), and in the SMC it is
$\sim\frac{1}{10}$ of that in the Solar neighbourhood (van den Bergh 1968; van
Genderen 1970; Lequeux et al.\ 1982; Bouchet et al.\ 1985). Dust is being
destroyed in the ISM by shocks and gas accretion, especially in Star Formation
Regions. Hence the dust-to-gas ratios in the ISM may only pose a lower limit
to the dust-to-gas ratios in the CSEs of obscured AGB stars.

Initial metallicities for relatively young ($10^{7-8}$ yr) field stars are
found to be about 2 and 5 times lower in the LMC and SMC, respectively,
compared to Solar metallicity (Spite et al.\ 1989a,b; Russell \& Bessell 1989;
Meliani et al.\ 1995; Luck et al.\ 1998). When depicting the chemical
evolution of the MCs, however, it is clear that the initial metallicity with
which stars were born a few Gyr ago was considerably lower --- roughly twice
--- than what is measured in massive stars today (de Freitas Pacheco et al.\
1998; Da Costa \& Hatzidimitriou 1998; Bica et al.\ 1998). Obscured AGB stars,
with a characteristic age of $10^9$ yr, are thus expected to have initial
metallicities somewhat lower than those measured in young field stars. Stars
in the central regions of the Milky Way galaxy are found to cover a range in
metallicity from sub- to super-solar, with the most easily observed obscured
AGB population likely to comprise relatively massive stars of slightly
super-solar initial metallicity (Rich 1988; Wood et al.\ 1998).

These considerations lead us to adopt a typical initial metallicity
(logarithmic) of $[Z/H] = -0.85\pm0.15$ for obscured AGB stars in the SMC,
$[Z/H] = -0.45\pm0.15$ in the LMC, $[Z/H] = 0.00\pm0.10$ in the Solar
neighbourhood, and $[Z/H] = 0.20\pm0.20$ in the Galactic Centre. The margins
are rough estimates for the typical range in initial metallicities, meaning
that the typical metallicity for the sample is not expected to lie outside of
these margins.

\subsection{Mass-loss rates and dust-to-gas ratios}

%
%
\begin{figure}[tb]
\centerline{\psfig{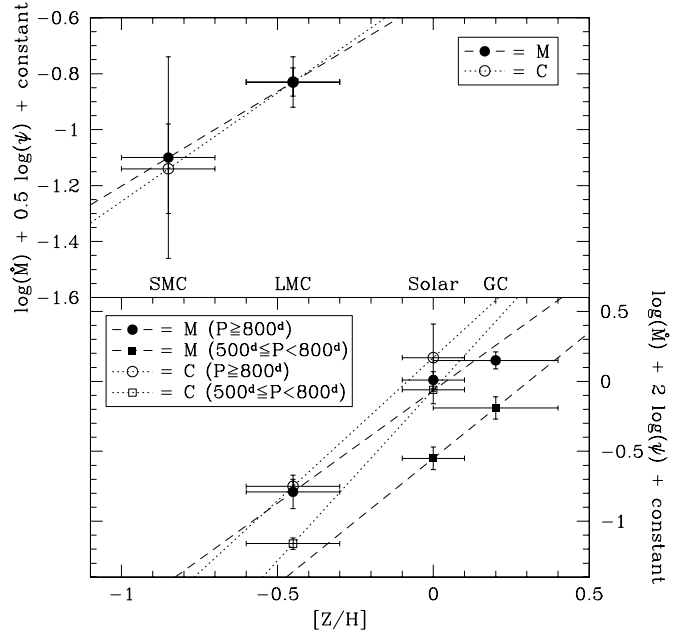}}
\caption[]{Mass-loss rates and dust-to-gas ratios for the obscured AGB stars
in the SMC, LMC, ``Solar neighbourhood'' and Galactic Centre, as a function of
their initial metallicities.}
\end{figure}

The values for the combination of mass-loss rates and dust-to-gas ratios as
derived from using Eqs.\ (4) \& (5) and listed in Tables 1 \& 2 are plotted in
Fig.\ 7. First order polynomials are drawn through the points. For the
obscured M-type stars with $500{\leq}P<800$ d the polynomial was forced to
cross the LMC data point, while the average was taken of the slopes obtained
by either omitting the ``Solar neighbourhood'' data point or the Galactic
Centre data point. The slopes of the polynomials can be used to constrain the
dependence of the mass-loss rate on initial metallicity $z$ (with $\log{z} =
[Z/H]$) and the dependence of the dust-to-gas ratio on $z$, assuming that
$\dot{M} \propto z^a$ and $\psi \propto z^b$. It then follows that the slope
for $(\log{\dot{M}} + 0.5 \log{\psi} + {\rm constant})$ equals $a + 0.5 b$,
and the slope for $(\log{\dot{M}} + 2 \log{\psi} + {\rm constant})$ equals $a
+ 2 b$. The sub-samples that had been selected according to the pulsation
periods have, in fact, very similar slopes, and hence their averages are
taken. Thus I find for the obscured M-type AGB stars that $a + 0.5 b \sim
0.68$ and $a + 2 b \sim 1.71$, whilst I find for the obscured carbon stars
that $a + 0.5 b \sim 0.78$ and $a + 2 b \sim 2.24$. This yields for the
dependences of the mass-loss rate and dust-to-gas ratio on initial metallicity
for obscured AGB stars:\\
\begin{tabular}{llll}
                       &   &                     &                     \\
$\log \dot{M}_{\rm M}$ & = & $0.3^{+0.7}_{-1.3}$ & $\log z$ + constant \\
$\log \dot{M}_{\rm C}$ & = & $0.3^{+0.6}_{-0.6}$ & $\log z$ + constant \\
$\log \psi_{\rm    M}$ & = & $0.7^{+0.6}_{-0.3}$ & $\log z$ + constant \\
$\log \psi_{\rm    C}$ & = & $1.0^{+0.3}_{-0.3}$ & $\log z$ + constant \\
                       &   &                     &                     \\
\end{tabular}\\
The reader should realise that these exponents are indicative, but not very
accurate. The errors (excluding those resulting from uncertainties in the
adopted initial metallicities) mainly result from the small number of
spectroscopically confirmed carbon and M-type stars in the SMC. If the initial
metallicity dependence of the mass-loss rate and dust-to-gas ratio is really
as similar as suggested here, then the obscured AGB stars in the SMC and LMC
may be compared without distinction by chemical type, and the resulting errors
on the exponents will become much smaller without significantly changing the
values of the exponents themselves. Clearly, it is shown that the method
employed here has the prospect of constraining the initial metallicity
dependences of the mass-loss rate and dust-to-gas ratio for obscured AGB
stars. Suitable data have only recently become available, and more such data
are needed to improve on the preliminary results.

\subsection{Obscured M-type AGB stars and carbon stars}

Obscured M-type AGB stars and obscured carbon stars have been treated
separately for two reasons: (i) the optical properties of their circumstellar
dust are different, possibly leading to differences in the constants of
proportionality $\alpha$ and $\beta$ in Eqs.\ (1) and (2), respectively, and
(ii) their dust-to-gas ratios and/or mass-loss rates may depend differently on
initial metallicity. I here discuss these two issues in more detail.

Both $\alpha$ and $\beta$ depend on the dust-type through the wavelength
dependent opacity $\kappa(\lambda)$ and the flux-weighted opacity $\kappa$,
respectively. Using Eq.\ (4) in the LMC yields indistinguishable distributions
of the obscured M-type AGB stars and of the obscured carbon stars. However, in
the LMC, obscured carbon stars are bolometrically fainter and exhibit lower
mass-loss rates than obscured M-type AGB stars (van Loon et al.\ 1999b), which
is expected to result in an offset between the distributions over
$(\log{\dot{M}} + 0.5 \log{\psi} + {\rm constant})$. The fact that this offset
is not seen must mean that, by mere coincidence, the differences in
$(\log{\dot{M}} + 0.5 \log{\psi})$ are counteracted upon by the differences in
$\kappa(\lambda) / \kappa$: at a given luminosity, dust-to-gas ratio and
mass-loss rate, obscured carbon stars are redder than obscured M-type AGB
stars. Also the coincidence between the distributions of obscured M-type AGB
stars and obscured carbon stars using Eq.\ (5) in the LMC for $500{\leq}P<800$
d must be coincidental and/or due to low number statistics (only two obscured
carbon stars), as the constant in Eq.\ (5) is proportional to
$\kappa^2(\lambda) \kappa$ and differences in optical properties between the
different dust species are likely to become apparent. This may explain at
least partly the differences in distributions between the obscured M-type AGB
stars and obscured carbon stars in the ``Solar neighbourhood''.

Obscured M-type AGB stars and obscured carbon stars seem to show very similar
dependencies of their dust-to-gas ratios and especially their mass-loss rates
on initial metallicity. There is no a-priori reason why the mass-loss
mechanism should have a different dependency on initial metallicity for stars
with different chemical types of circumstellar dust, other than due to a
different dependency of their dust-to-gas ratios on initial metallicity. The
similarity between the dependencies of the dust-to-gas ratio on initial
metallicity is surprising. For obscured M-type AGB stars one may expect that
the fraction of metals that are available for the formation of dust particles
scales directly with the oxygen abundance in the photosphere, which scales at
least approximately directly with the initial metallicity of the star. Hence a
direct proportionality between dust-to-gas ratio and initial metallicity may
not come as a surprise for obscured M-type AGB stars. For obscured carbon
stars, however, the situation is very different: carbon stars only become
carbon stars after $3^{\rm rd}$ dredge-up has enhanced the carbon abundance in
the photosphere from $[C/O]<1$ to $[C/O]>1$. For obscured carbon stars it is
crucial to know the photospheric abundances of both carbon and oxygen, because
the carbon is locked into CO molecules until oxygen exhaustion and hence only
the carbon excess is available for dust formation. It was thought that at
lower initial metallicity, the lower oxygen abundance would make it easier for
$3^{\rm rd}$ dredge-up to raise $[C/O]$ above unity, but the fact that no
optically bright luminous carbon stars were found in the MCs meant that it is
not that simple (Iben 1981). Not only is it poorly understood how $3^{\rm rd}$
dredge-up depends on initial metallicity (but see Marigo et al.\ 1999), there
is also a second important phenomenon active: carbon star formation is avoided
as long as the stellar mantle is massive enough to yield pressures and
temperatures at the bottom of the convective layer sufficiently high for
processing of carbon into oxygen and nitrogen to occur (Hot Bottom Burning;
Iben \& Renzini 1983; Wood et al.\ 1983). Thus it remains to be seen how the
carbon excess for obscured carbon stars depends on initial metallicity (see
also van Loon et al.\ 1999a). The data and analysis presented here suggest
that the carbon excess for obscured carbon stars may be directly proportional
to the initial metallicity.

\section{Summary}

%
%
\begin{table*}
\caption[]{Near-IR stars near IRAS point sources in the direction of the LMC
(LI=LI-LMC: Schwering \& Israel 1990) that are candidate obscured AGB stars.
Listed are IRAS flux densities (in Jy), near-IR position, distance to the IRAS
source (in arcsec), near-IR magnitudes, and bolometric magnitude assuming
association (see text). Values between parentheses are 1-$\sigma$ errors.}
\begin{tabular}{rcccllllcccl}
\hline\hline
LI & $F_{12}$ & $F_{25}$ & $F_{60}$ &
$\alpha$(2000) & $\delta$(2000) & $\Delta$ &
$J\pm\sigma_J$ & $H\pm\sigma_H$ &
$K\pm\sigma_K$ & $L\pm\sigma_L$ &
$M_{\rm bol}$ \\
\hline
        203 & 0.31 & 0.20 &       &
4 55 40.6 & $-69$ 26 40 &  8 &
\llap{1}1.955(0.017) &                      &
\llap{1}1.858(0.039) &                      &
$-5.8$ \\
            &      &      &       &
4 55 41.7 & $-69$ 26 22 & 26 &
        8.776(0.016) &                      &
        7.719(0.021) &                      &
$-8.3$ \\
            &      &      &       &
4 55 35.7 & $-69$ 26 56 & 22 &
        8.455(0.017) &                      &
        7.346(0.022) &         6.850(0.073) &
$-8.6$ \\
        987 & 0.48 & 0.43 &       &
5 24 41.7 & $-69$ 15 20 & 15 &
\llap{1}3.006(0.069) & \llap{1}1.906(0.035) &
\llap{1}1.622(0.053) &                      &
$-5.7$ \\
\llap{1}284 & 0.37 & 0.79 &      &
5 32 38.9 & $-68$ 25 22 & 13 &
                     & \llap{1}4.015(0.049) &
\llap{1}1.896(0.023) &         9.127(0.071) &
$-5.8$ \\
\llap{1}522 & 1.00 & 0.92 &      &
5 40 13.1 & $-69$ 22 50 &  3 &
\llap{1}0.660(0.015) &         9.919(0.018) &
        8.612(0.017) &         6.886(0.032) &
$-7.4$ \\
\llap{1}795 & 0.25 & 0.28 & 0.16 &
5 56 42.7 & $-67$ 53 21 & 29 &
\llap{1}2.058(0.025) & \llap{1}1.157(0.019) &
\llap{1}0.907(0.024) & \llap{1}0.336(0.129) &
$-5.7$ \\
\hline
\end{tabular}
\end{table*}

In conclusion, the comparison between the dust-to-gas ratios and mass-loss
rates of obscured AGB stars in the SMC, LMC, and Milky Way suggests that the
dust-to-gas ratios in the outflows of obscured AGB stars depend approximately
linearly on the initial metallicity. This was suspected by Habing et al.\
(1994) but still awaited observational support. The total (gas+dust) mass-loss
rates of obscured AGB stars show a much weaker dependence on initial
metallicity. No appreciable differences are found between these relations for
obscured M-type AGB stars and obscured carbon stars.

\acknowledgements{I am greatly indebted to Dr.\ Albert Zijlstra for the
motivation to present this study, and to Prof.Dr.\ Teije de Jong for his
interest and stimulating discussions. I would like to thank Drs.\ Martin
Groenewegen, Rens Waters, and Joana Oliveira for reading an earlier version of
the manuscript, and Fred Marang for the excellent support during the near-IR
observations at Sutherland. I also thank Prof.Dr.\ Harm Habing and Dr.\ Peter
Wood for sharing their results prior to publication, Dr.\ Martin Groenewegen
for providing his published data in electronic form, and an anonymous referee
whose recommendations lead to considerable improvements in the paper. A large
part of this work was done while JvL was a student at ESO and the University
of Amsterdam. A Joaninha est\'{a} sempre nas coisas mais importantes da vida.}

\appendix

\section{New near-IR counterparts of IRAS sources in the LMC}

Periods of weather conditions that were too poor for long-term photometric
monitoring at the South African Astronomical Observatory (SAAO) at Sutherland,
South Africa, in December 1997 were used to search for near-IR counterparts
of IRAS point sources in the direction of the LMC. This was done on the 1.9 m
telescope with the Mk~{\sc iii} scanning photometer in the K-band. An aperture
of $12^{\prime\prime}$ was used, chopping and nodding with a throw of
$30^{\prime\prime}$. The search was limited to objects brighter than $K=13$
mag. The areas around five IRAS point sources suspected to be obscured AGB
stars (paper I) were searched. The candidate near-IR counterparts found are
listed in Table A1, where the photometry is in the SAAO system (Carter 1990),
i.e.\ the J-band magnitude is transformed to the 0.75 m telescope system. One
object was re-observed under good photometric conditions (LI-LMC1284),
together with the star HR2015 ($\delta$ Dor) for photometric calibration.
Positions have been estimated by comparing the position of the diaphragm in
the (red) acquisition video images with the second generation Digital Sky
Survey, and are accurate to $\sim2^{\prime\prime}$.

I retrieved 12, 25 and 60 $\mu$m data from the IRAS data base server in
Groningen\footnote{The IRAS data base server of the Space Research
Organisation of the Netherlands (SRON) and the Dutch Expertise Centre for
Astronomical Data Processing is funded by the Netherlands Organisation for
Scientific Research (NWO). The IRAS data base server project was also partly
funded through the Air Force Office of Scientific Research, grants AFOSR
86-0140 and AFOSR 89-0320.} (Assendorp et al.\ 1995). Point sources were
recovered by means of $2\times2$ square degree maps with $0.5^\prime$ pixels.
The flux density was measured from a trace through the position of the source
using the command {\sc scanaid} in the Groningen {\sc gipsy} data analysis
software. LI-LMC203 shows a hint of duplicity: two similarly bright sources
separated by one arcminute. LI-LMC987 looks slightly extended, and LI-LMC1284
is on top of brighter emission. LI-LMC1522 and especially LI-LMC1795 are
isolated. Assuming identification of near-IR and IRAS sources, the spectral
energy distribution were integrated graphically to yield bolometric
magnitudes.

LI-LMC203 is not identified with certainty. The best spatial coincidence is
for the first listed in Table A1, that has blue near-IR colours incompatible
with mass-losing AGB stars. There are two much brighter near-IR sources with
moderately red $(J-K)$ nearby, of which the third listed in Table A1 is a
cluster of $\sim10$ stars within a diameter of $\sim12^{\prime\prime}$. The
proposed near-IR counterparts of LI-LMC987 and 1795 have near-IR colours
consistent with red giants without mass loss and are probably not associated
with the IRAS sources. The near-IR counterpart of LI-LMC1284 is a heavily
obscured AGB star, with IR colours compatible with either carbon or
oxygen-rich dust. LI-LMC1522 is also identified as a dust-enshrouded star,
with IR colours suggesting oxygen-rich dust.

\section{Expansion velocities}

Here the expansion velocities of AGB stars are briefly discussed in order to
arrive at a justified calibration of the $v_{\rm exp}(L)$ scaling relation
(Eq.\ (3)) for LMC stars. I consider $v_{\rm exp}$ derived from the separation
of the peaks of OH maser line profiles, and from the width of CO(1--0)
emission for Milky Way stars without OH measurements. The latter are divided
by 1.12, following Groenewegen et al.\ (1998; see also Lewis 1991).

%
%
\begin{figure*}[tb]
\centerline{\psfig{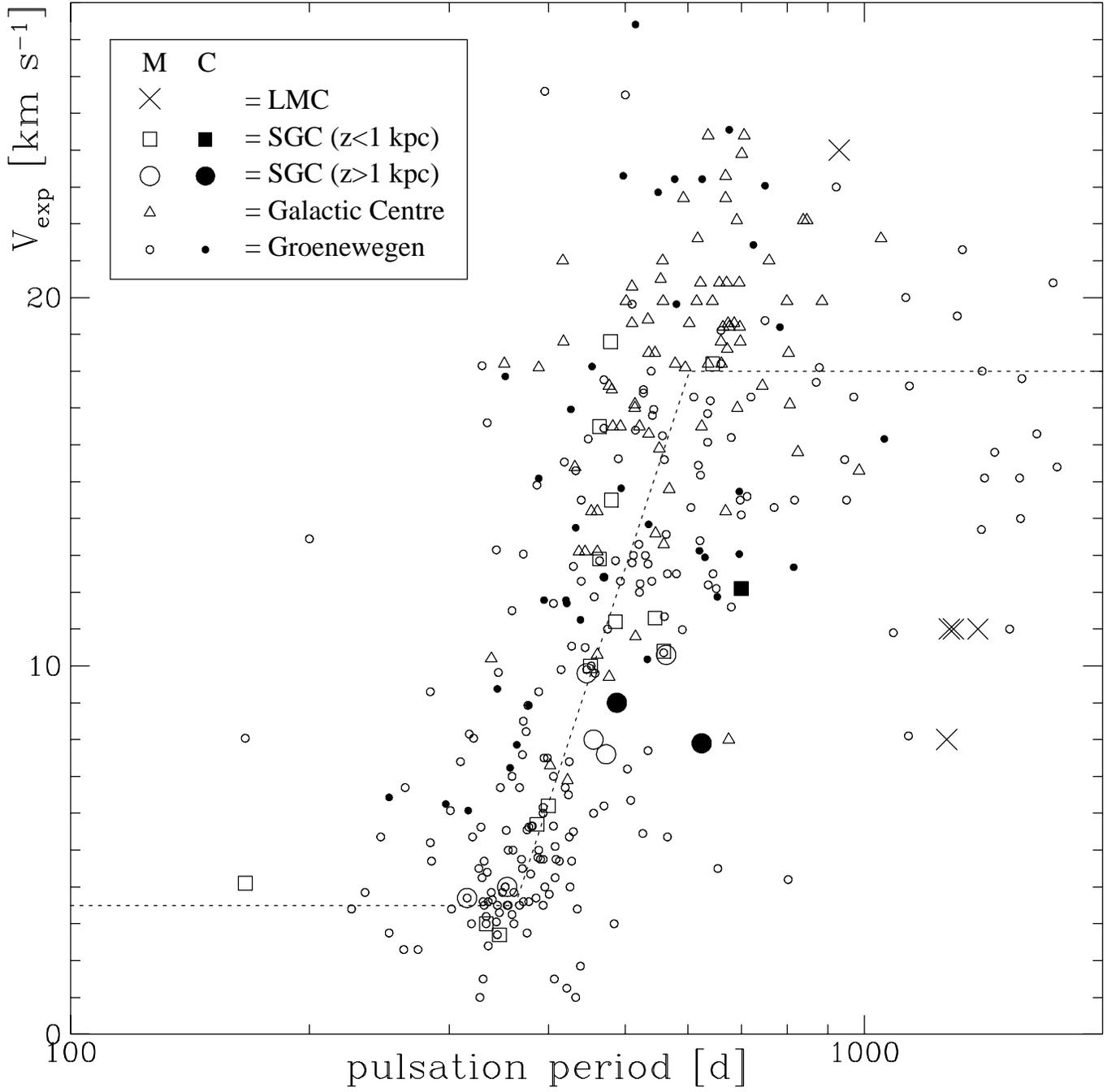}}
\caption[]{Expansion velocity versus pulsation period, for samples in the LMC
(Wood et al.\ 1992; Wood 1998; van Loon et al.\ 1998b), South Galactic Cap
(Whitelock et al.\ 1994), Galactic Centre (Wood et al.\ 1998) and ``Solar
neighbourhood'' (Groenewegen et al.\ 1998). Carbon stars are represented by
filled symbols. The dotted line is drawn to guide the eye: expansion
velocities are low for $P\lsim350$, increase for longer periods, and reach a
more constant level for $P\gsim600$. Low-metallicity stars in the LMC and SGC
samples (at more than 1 kpc from the galactic plane) have lower $v_{\rm exp}$
than stars with $\gsim$solar metallicity.}
\end{figure*}

The $v_{\rm exp}$ is plotted against $P$ in Fig.\ B1. Periods for the LMC
stars are from Wood et al.\ (1992) and Wood (1998), and their $v_{\rm exp}$
are from Wood et al.\ (1992) and van Loon et al.\ (1998b). Carbon stars
(filled symbols) are distinguished from oxygen-rich, M-type stars (open
symbols). As an AGB star evolves, $P$ increases. Mass loss becomes important
for $P\gsim400$ d and reaches a maximum for $P\gsim600$ d (cf.\ Jura 1986).
This evolution is also reflected in $v_{\rm exp}$, which increases when
$P\gsim400$ d and levels off at a value $v_{\rm exp}\sim18$ km s$^{-1}$ for
$P\gsim600$ d (dotted line in Fig.\ B1, see also Lewis 1991). The individual
data sets (and Lewis 1991) suggest that AGB stars with $P\lsim350$ d evolve at
$\sim$constant $v_{\rm exp}$ typically 3 to 4 km s$^{-1}$. This is also seen
in (intrinsic) S-type stars with semi-regular variability (Fig.\ 18 in
Jorissen \& Knapp 1998). This may be a slow wind supported by another
mechanism, such as radiation pressure on molecules (cf.\ Sahai \& Liechti
1995; Steffen et al.\ 1997, 1998). It agrees with the observation that SiO
masers, that depend on shocks by stellar pulsation (Alcolea et al.\ 1990), may
be present for $P\gsim350$ d and become ubiquitous when $P\gsim700$ d
(Izumiura et al.\ 1994).

Carbon stars appear to have somewhat larger $v_{\rm exp}$ than M-type stars at
the same $P$, although the (few) SGC carbon stars do not obey this trend. The
data is also consistent with smaller $P$ for carbon stars at the same $v_{\rm
exp}$.

SGC stars within 1 kpc from the galactic plane (squares) are distinguished
from SGC stars beyond that (large circles). The latter are presumably of
sub-solar initial metallicity and have smaller $v_{\rm exp}$ than the former.

Wood et al.\ (1992) showed that $v_{\rm exp}$ is smaller at LMC metallicity
than at solar metallicity, providing supportive evidence for Eq.\ (3). The LMC
star with $v_{\rm exp}=24$ km s$^{-1}$ is IRAS04553$-$6825, a very luminous
RSG (van Loon et al.\ 1998b, and references therein). The other LMC stars are
AGB stars with $v_{\rm exp}\sim10$ km s$^{-1}$. The OH/IR stars in the
Groenewegen sample with $P\gsim1000$ have $v_{\rm exp}\sim10$ to 20 km
s$^{-1}$, suggesting that initial metallicities of these Milky Way stars are
higher than the LMC stars. The expansion velocities of the obscured AGB stars
in the Galactic Centre range up to $v_{\rm exp}=25$ km s$^{-1}$, and initial
metallicities of two to three times solar have been suggested by Wood et al.\
(1998). Considering all this, Eq.\ (3) is calibrated by demanding a star with
LMC metallicity and $L=30,000$ L$_\odot$ ($M_{\rm bol}=-6.5$ mag) to have
$v_{\rm exp}=10$ km s$^{-1}$.


\begin{thebibliography}{}
\bibitem[1990]{}
Alcolea J., Bujarrabal V., G\'{o}mez-Gonz\'{a}lez J., 1990, A\&A 231, 431
\bibitem[1995]{}
Assendorp R., Bontekoe T.R., de Jonge A.R.W., et al., 1995, A\&AS 110, 395
\bibitem[1998]{}
Bedding T.R., Zijlstra A.A., 1998, ApJ 506, L47
\bibitem[1998]{}
Bica E., Geisler D., Dottori H., et al., 1998, AJ 116, 723
\bibitem[1998]{}
Blommaert J.A.D.L., van der Veen W.E.C.J., van Langevelde H.J., Habing H.J.,
Sjouwerman L.O., 1998, A\&A 329, 991
\bibitem[1985]{}
Bouchet P., Lequeux J., Maurice E., Pr\'{e}vot L., Pr\'{e}vot-Burnichon M.-L.,
1985, A\&A 149, 330
\bibitem[1990]{}
Carter B.S., 1990, MNRAS 242, 1
\bibitem[1985]{}
Clayton G.C., Martin P.G., 1985, ApJ 288, 558
\bibitem[1998]{}
Da Costa G.S., Hatzidimitriou D., 1998, AJ 115, 1934
\bibitem[1998]{}
de Freitas Pacheco J.A., Barbuy B., Idiart T., 1998, A\&A 332, 19
\bibitem[1986]{}
Gail H.-P., Sedlmayr E., 1986, A\&A 161, 201
\bibitem[1998]{}
Groenewegen M.A.T., Blommaert J.A.D.L., 1998, A\&A 332, 25
\bibitem[1998]{}
Groenewegen M.A.T., Whitelock P.A., Smith C.H., Kerschbaum F., 1998, MNRAS
293, 18
\bibitem[1994]{}
Habing H.J., Tignon J., Tielens A.G.G.M., 1994, A\&A 286, 523
\bibitem[1981]{}
Iben I., 1981, ApJ 246, 278
\bibitem[1983]{}
Iben I., Renzini A., 1983, ARA\&A 21, 271
\bibitem[1994]{}
Izumiura H., Deguchi S., Hashimoto O., et al., 1994, ApJ 437, 419
\bibitem[1998]{}
Jorissen A., Knapp G.R., 1998, A\&AS 129, 363
\bibitem[1986]{}
Jura M., 1986, ApJ 303, 327
\bibitem[1982]{}
Koornneef J., 1982, A\&A 107, 247
\bibitem[1999]{}
Lamers H.J.G.L.M., Cassinelli J.P., 1999, in: Introduction to Stellar Winds.
Cambridge University Press, ch. 7
\bibitem[1982]{}
Lequeux J., Maurice E., Pr\'{e}vot-Burnichon M.-L., et al., 1982, A\&A 113,
L15
\bibitem[1991]{}
Lewis B.M., 1991, 101, 254
\bibitem[1997]{}
Loup C., Zijlstra A.A., Waters L.B.F.M., Groenewegen M.A.T., 1997, A\&AS 125,
419 (paper I)
\bibitem[1999]{}
Loup C., Josselin E., Cioni M.-R., et al., 1999, in: Asymptotic Giant Branch
Stars (IAU Symposium no.\ 191), eds.\ C. Waelkens, T. Lebertre, A. L\`{e}bre.
ASP Conf. Ser., p561
\bibitem[1998]{}
Luck R.E., Moffett T.J., Barnes T.G., Gieren W.P., 1998, AJ 115, 605
\bibitem[1999]{}
Marigo P., Girardi L., Bressan A., 1999, A\&A 344, 123
\bibitem[1995]{}
Meliani M.T., Barbuy B., Richtler T., 1995, A\&A 304, 347
\bibitem[1988]{}
Rich R.M., 1988, AJ 95, 828
\bibitem[1989]{}
Russell S.C., Bessell M.S., 1989, ApJS 70, 865
\bibitem[1995]{}
Sahai R., Liechti S., 1995, A\&A 293, 198
\bibitem[1990]{}
Schwering P.B.W., Israel F.P., 1990, A catalog of IRAS sources in the
Magellanic Clouds. Kluwer, Dordrecht
\bibitem[1989a]{}
Spite F., Spite M., Fran\c{c}ois P., 1989a, A\&A 210, 25
\bibitem[1989b]{}
Spite M., Spite F., Barbuy B., 1989b, A\&A 222, 35
\bibitem[1997]{}
Steffen M., Szczerba R., Men'shchikov A., Sch\"{o}nberner D., 1997, A\&AS 126,
39
\bibitem[1998]{}
Steffen M., Szczerba R., Sch\"{o}nberner D., 1998, A\&A 337, 149
\bibitem[1968]{}
van den Bergh S., 1968, J. R. Astron. Soc. Can. 62, 145 \& 178
\bibitem[1970]{}
van Genderen A.M., 1970, A\&A 7, 49
\bibitem[1997]{}
van Loon J.Th., Zijlstra A.A., Whitelock P.A., et al., 1997, A\&A 325, 585
(paper III)
\bibitem[1998a]{}
van Loon J.Th., Zijlstra A.A., Whitelock P.A., et al., 1998a, A\&A 329, 169
(paper IV)
\bibitem[1998b]{}
van Loon J.Th., te Lintel Hekkert P., Bujarrabal V., Zijlstra A.A., Nyman,
L.-\AA., 1998b, A\&A 337, 141
\bibitem[1999a]{}
van Loon J.Th., Zijlstra A.A., Groenewegen M.A.T., 1999a, A\&A 346, 805
\bibitem[1999b]{}
van Loon J.Th., Groenewegen M.A.T., de Koter A., et al., 1999b, A\&A 351, 559
\bibitem[1999]{}
Walker A.R., 1999, in: Post Hipparcos Cosmic Candles, eds.\ A. Heck \& F.
Caputo. Astrophysics and Space Science Library vol.\ 237, Kluwer, Dordrecht,
p125
\bibitem[1994]{}
Whitelock P.A., Menzies J.W., Feast M.W., et al., 1994, MNRAS 267, 711
\bibitem[1998]{}
Wood P.R., 1998, A\&A 338, 592
\bibitem[1999]{}
Wood P.R., 1999, in: Asymptotic Giant Branch Stars (IAU Symposium no.\ 191),
eds.\ C. Waelkens, T. Lebertre, A. L\`{e}bre. ASP Conf. Ser., p151
\bibitem[1983]{}
Wood P.R., Bessell M.S., Fox M.W., 1983, ApJ 272, 99
\bibitem[1992]{}
Wood P.R., Whiteoak J.B., Hughes S.M.G., et al., 1992, ApJ 397, 552
\bibitem[1998]{}
Wood P.R., Habing H.J., McGregor P.J., 1998, A\&A 336, 925
\bibitem[1996]{}
Zijlstra A.A., Loup C., Waters L.B.F.M., et al., 1996, MNRAS 279, 32 (paper
II)
\end{thebibliography}
\end{document}